\documentclass[review]{elsarticle}

\journal{Nuclear Instruments and Methods in Physics Research A}

\usepackage{amssymb}

\usepackage{amsthm}
\usepackage{physics}
\usepackage{todonotes}
\usepackage{ulem}
\usepackage{xspace}

\usepackage{lineno}

\usepackage{cleveref}
\usepackage{url}
\usepackage{multirow}
\usepackage{multicol}
\usepackage{subcaption}

\begin{document}

\begin{frontmatter}

\title{Study of  Silicon Photomultiplier Performance at Different Temperatures}

\author[jinraddress]{N.~Anfimov\corref{cor1}}
\ead{anphimov@jinr.ru}
\author[jinraddress]{D.~Fedoseev}
\author[jinraddress]{A.~Rybnikov}
\author[jinraddress]{A.~Selyunin\corref{cor1}}
\ead{selalsebog@gmail.com}
\author[jinraddress]{S.~Sokolov}
\author[jinraddress]{A.~Sotnikov}
\cortext[cor1]{Corresponding authors}
\address[jinraddress]{Joint Institute for Nuclear Research, Joliot-Curie 6, Dubna, Russia, 141980}

\begin{abstract}
Decreasing the operation temperature of a Silicon Photo-Multiplier (SiPM) leads to a drop in its dark noise. Some experiments consider cold temperatures as an option for low noise applications of SiPM. One of those is the TAO detector, which requires operation at $T\approx -50~^\circ$C. A significant dependence of the Photon Detection Efficiency (PDE) of a SiPM on different temperatures was reported with a drastic drop around this temperature. In this paper, we present studies of performance for two samples of SiPMs from Hamamatsu and AdvanSID~(FBK) companies in a broad temperature range. No significant difference for the PDE was observed. 
\end{abstract}


\end{frontmatter}

\section{Introduction}
\label{sec:introduction}
A Silicon photomultiplier (SiPM) is an array of small avalanche photodiodes, called pixels, that operate above the breakdown voltage (overvoltage (OV)) in a Geiger mode. The breakdown voltage is the minimum reverse bias voltage that leads to self-sustaining avalanche multiplication in every SiPM's pixel. Such a structure allows for detecting and working with very low-intensity light in a proportional mode with a range limited by the total number of pixels. SiPMs are being introduced to many different applications. Nowadays they are widely used in experimental physics and medical diagnostic techniques. The main advantages with respect to conventional photomultipliers are higher detection efficiency, less sensitivity to magnetic fields, compact dimensions, and gain of up to 10$^6$. On the other hand, there are drawbacks of SiPMs: high dark rate, optical crosstalk and temperature dependence of some parameters. 

\begin{figure}[!h]
   \centering
  \includegraphics[width=230px]{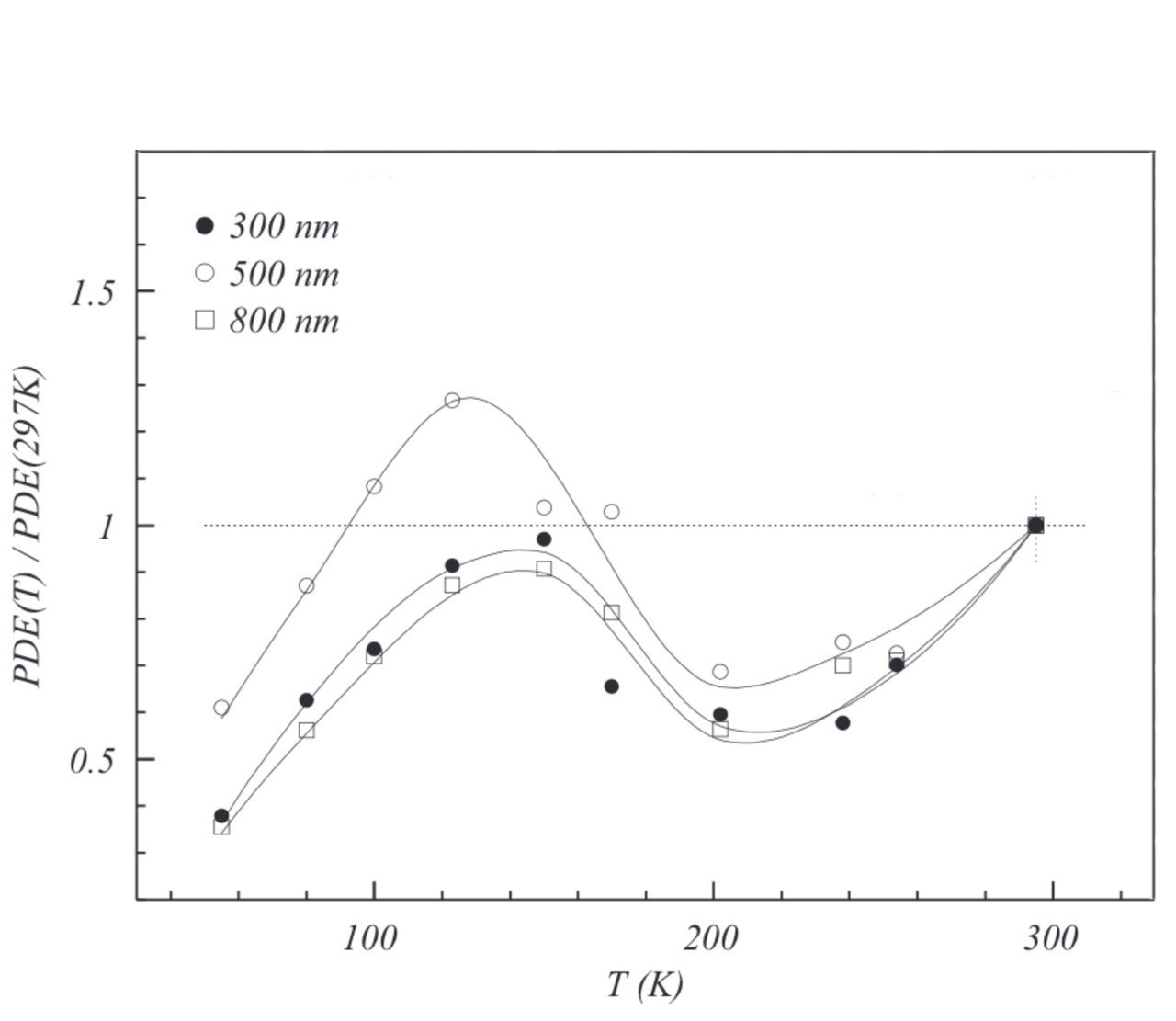} \\ 
   \caption{Normalized photon detection efficiency as a function of temperature for fixed overvoltage of 2V and three wavelengths (300 nm, 500 nm, and 800 nm) for 1$\times$1 mm$^2$ FBK-IRST SiPM type T6-V1-PD~\cite{italian}.}
  \label{fig:fig1}
\end{figure}

Currently, a satellite experiment for the JUNO neutrino experiment called TAO is under consideration\cite{tao1}. It is proposed that TAO will precisely measure the primary neutrino spectrum from the nuclear reactors. The TAO detector is a spherical vessel filled with a liquid scintillator covered by photosensors. To achieve the required energy resolution of less than  $2~\%$ at 1 MeV for the TAO experiment one needs to maximize the light detection efficiency and minimize the noise level. SiPM is a good candidate as a photosensor for TAO since it has a higher photon detection efficiency(PDE) compared to conventional PMTs. To reduce the high noise level of the SiPM it is proposed to operate the TAO detector at low temperature $T\approx -50~^\circ$C~\cite{tao2}. But there is a relevant study \cite{italian} where a significant drop of the PDE for 1$\times$1 mm$^2$ FBK-IRST SiPM type T6-V1-PD  at $T\approx-50~^\circ$C$~(\approx220~$K) was observed (see \cref{fig:fig1}). Such a drop may totally wash out the advantage of SiPM usage. Assuming the effect might exist for other SiPM models we have studied two SiPM samples: Hamamatsu MPPC-S13360-6025CS~\cite{ham} (single element), AdvanSID ASD-NUV4S-P-4x4TD~\cite{FBK} (SiPM array). 

The paper is organized as follows. In \cref{sec:setup} we describe experimental setup and methods that we used in our study. Next \cref{sec:results} presents results for both SiPM samples at different temperatures. And finally, in  \cref{sec:summary} we draw our conclusions.

\section{Experimental setup}
\label{sec:setup}

In our measurements, a Dewar's vessel is used as a cryostat to keep Liquid Nitrogen (LN). Nitrogen vapor produces a gradient of temperature at different heights inside the Dewar. By moving the SiPM one can provide different temperatures in a broad range from room conditions down to LN environment. To prevent any light leakage into the Dewar's vessel it is placed inside a dark room that is completely insulated from exterior light.

\subsection{Incident light stability}

To provide a valid study one needs to guarantee that light intensity is stable in time and with environmental changes. In our measurements, we use a stabilized LED light source with $\lambda=$ 425~nm wavelength produced by the HVSys company~\cite{HV_sys}. The light from the LED is transported to the SiPM by means of a plastic optical fiber. The light traveling through the optical fiber may vary at level of about 10~\% with temperature changes. It might be driven by changing the optical properties between the fiber's core and the cladding. To reduce such a behavior, temperature stabilization is implemented along the fiber's length. 

\begin{figure}[htbp]
  \begin{minipage}[ht]{0.29\linewidth}
  \label{fig:fig_setup_a}
  \centering
   \subcaptionbox{}
          {\includegraphics[width=\linewidth]{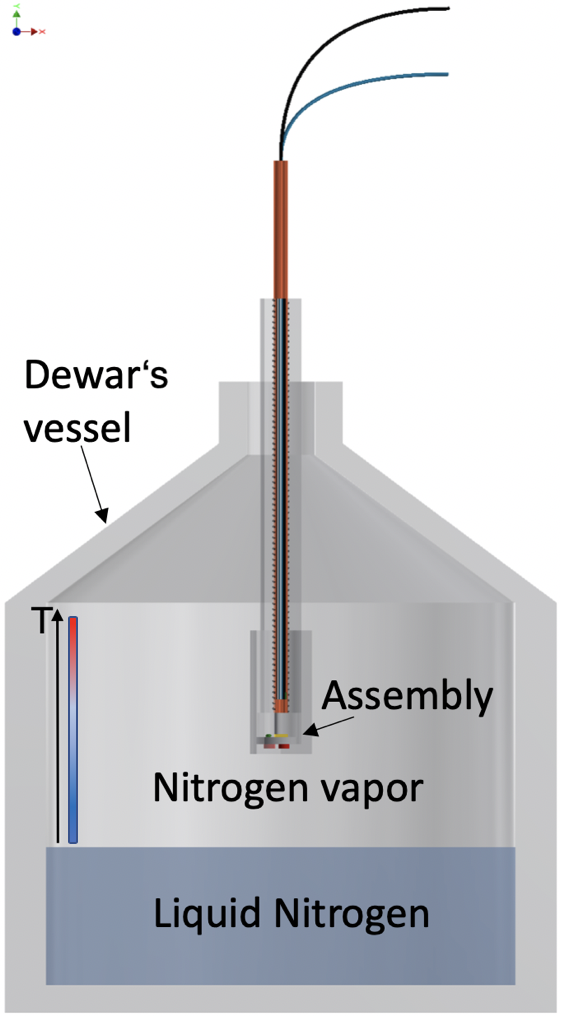}}
  \end{minipage}
  \hfill 
  \begin{minipage}[ht]{0.69\linewidth}
    \label{fig:fig_setup_b}
  \centering
   \subcaptionbox{}
          {\includegraphics[width=\linewidth]{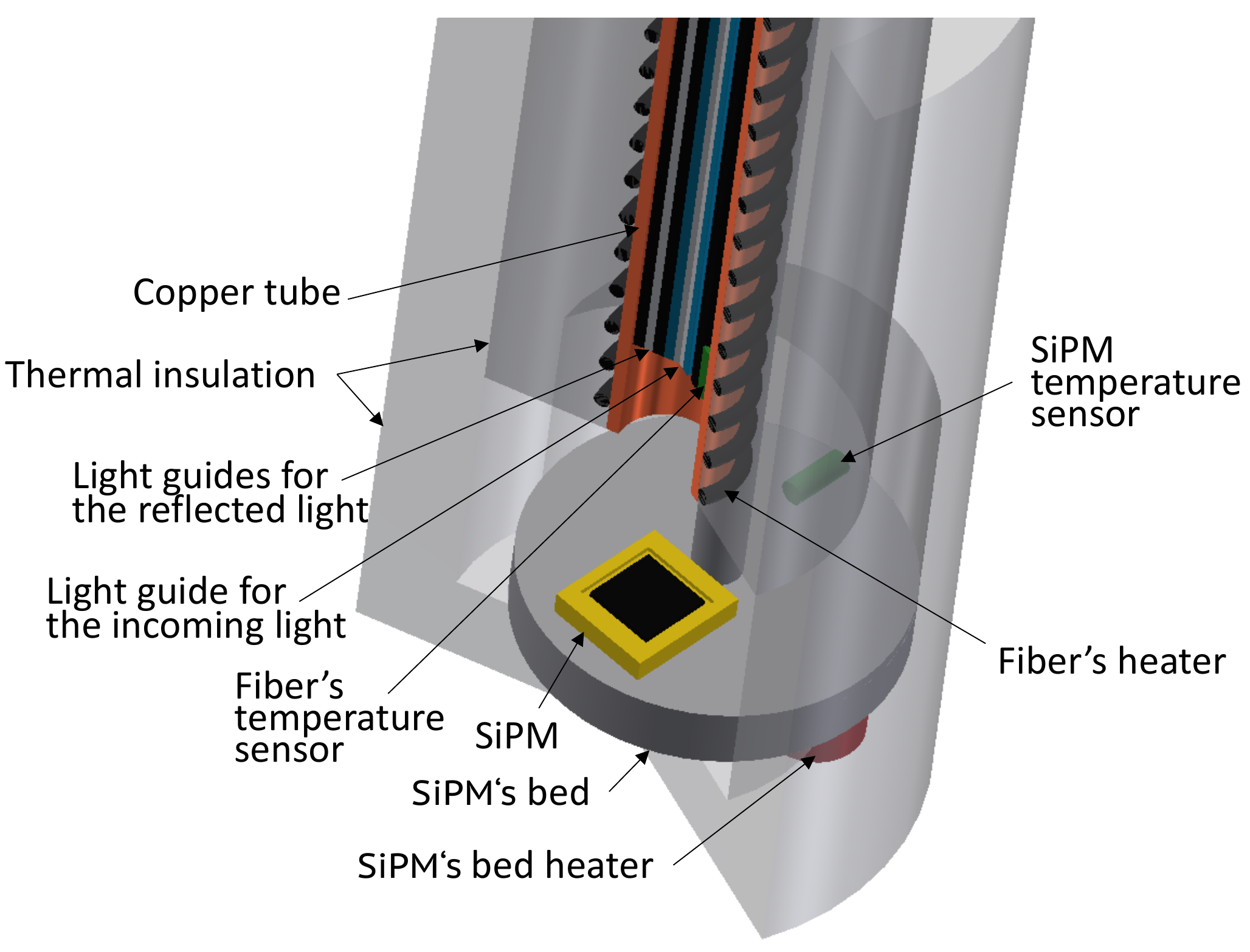}}
   
  \end{minipage}
  \hfill
  \caption{Schematic view of the experimental setup: a - Dipped probe into the Dewar's vessel, b - Probe schematic with SiPM's bed and light guides  are mounted in a copper tube.}
  \label{fig:fig_setup}  
\end{figure}

\subsection{Setup description}
The cryostat vessel of $\sim$30 liters is filled to about 1/3 with Liquid Nitrogen (see \cref{fig:fig_setup}a). An assembly of a light delivery system  (see \cref{fig:fig_setup}b) with the SiPM is moving along the cryostat depth. The light delivery system is an optical fiber bundle of 6 fibers which is placed inside a copper pipe that screens it from temperature changes. The light is sent to the SiPM or reflective plate (calibration) through the central fiber. Then the light reflects back from the SiPM and travels to the other fibers which are used to monitor light stability with temperature changes. The pipe is wound by a heating cable with feedback provided by a thermal sensor that is placed inside the pipe. The heating cable keeps the fiber bundle at a stable temperature. The assembly is additionally thermally insulated on the outside. 

A reflective plate that replaces the silicon dice (SiPM) for the light stability calibration is placed in front of the fiber bundle in order to reflect the light back through the other 5 fibers. In this measurement, we use quite high light intensity which is enough to monitor the reflected light by means of a PIN-photodiode ThorLabs FDS100-CAL~\cite{thorlabs} and a Keithley 6487 Picoammeter/Voltage source. The measurement gives an estimation of the light stability for different environmental temperatures. Overall light stability in our system is reached at the level of 1\%.

\begin{figure}[!h]
   \centering
  \includegraphics[width=300px]{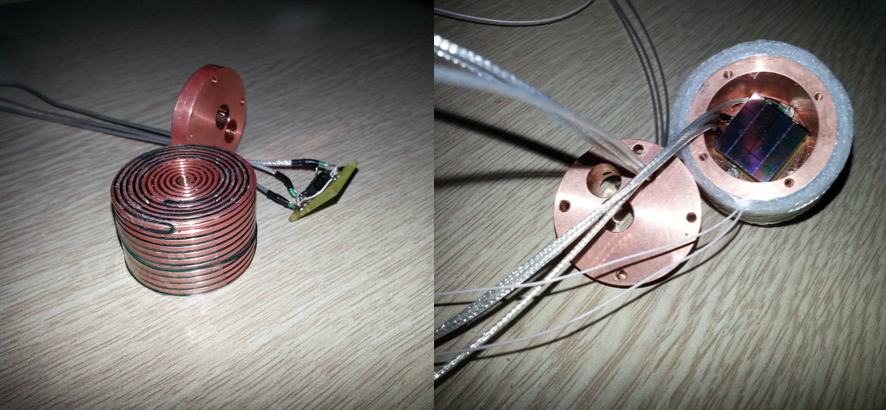} \\ 
   \caption{Copper tub is wound with heating cable and PCB with ASD-NUV4S-P-4x4TD}
  \label{fig:tub}
\end{figure}

The reflection plate is then replaced with a single SiPM which is stuck to the aluminum substrate (bed) with a thermal grease with an embedded thermosensor on its backside. For the AdvanSID SiPM array we use a copper tub. The SiPM and thermosensor are placed inside the tub which is wound with heating cable(see~\cref{fig:tub}). In both cases, the temperature stability is guaranteed with precision better than $1^\circ$C.

\subsection{Readout chain and data acquisition}
\begin{figure}[!h]
   \centering
  \includegraphics[width=270px]{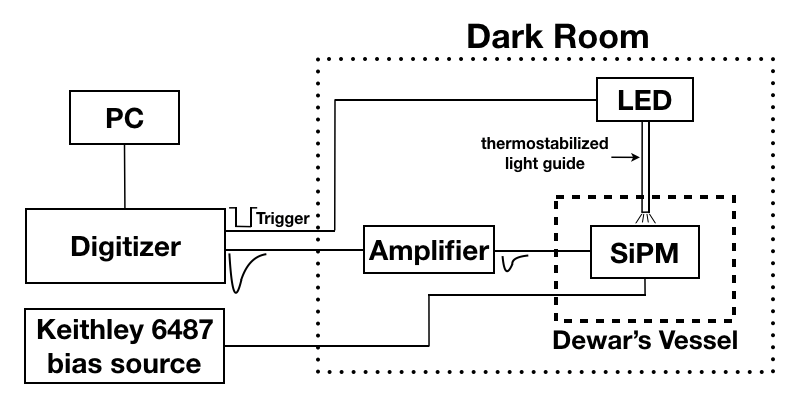} \\ 
   \caption{Readout chain circuit.}
  \label{fig:readout_chain}
\end{figure}

A readout chain in our measurements consists of a SiPM power supply source, an amplifier and a high-speed digitizer(see~\cref{fig:readout_chain}). Apart from the readout chain, two power supply sources are utilized to supply the heating cables. All of this equipment is located outside the darkroom, except the warm amplifier that sits close to the Dewar.  

To provide a bias voltage to the SiPM, the Keithley 6487 is used. The source automatically measures the current-voltage (I-V) curves of the SiPM that are used for a breakdown voltage determination.

A signal from the SiPM is digitized by means of DRS4 \cite{drs} evaluation board that can sample a signal within a range of (0.7-5)~GS/s with about 11.5 effective bits precision. It operates in external-trigger mode and is synchronized with the LED trigger signal. The saved data are analyzed by software using the ROOT package \cite{root}. We put a high-gain amplifier into the readout chain  to increase the sensitivity of the ADC by 225 times in order to distinguish single photoelectrons.

\subsection{Data analysis}

\begin{figure}[htbp]
  \begin{minipage}[ht]{\linewidth}
  \label{fig:fig_pde_1}
  \centering
   \subcaptionbox{}
          {\includegraphics[width=\linewidth]{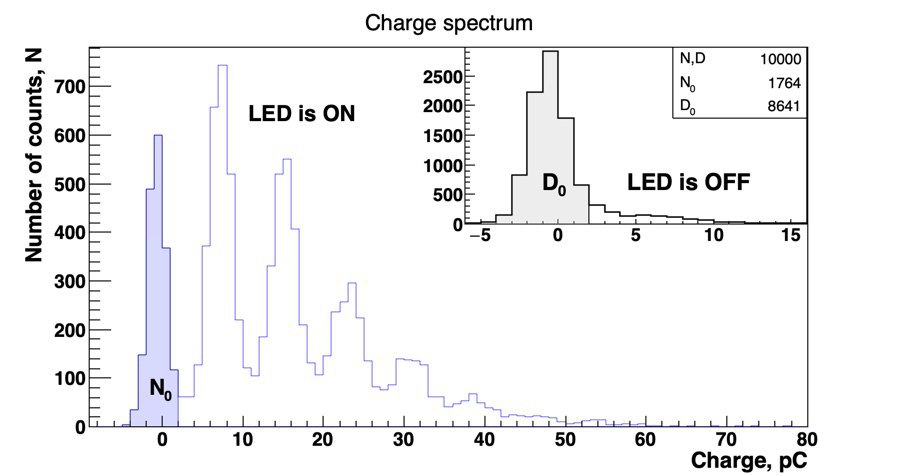}}
  \end{minipage}
  \hfill 
  \begin{minipage}[ht]{\linewidth}
    \label{fig:fig_pde_2}
  \centering
   \subcaptionbox{}
          {\includegraphics[width=\linewidth]{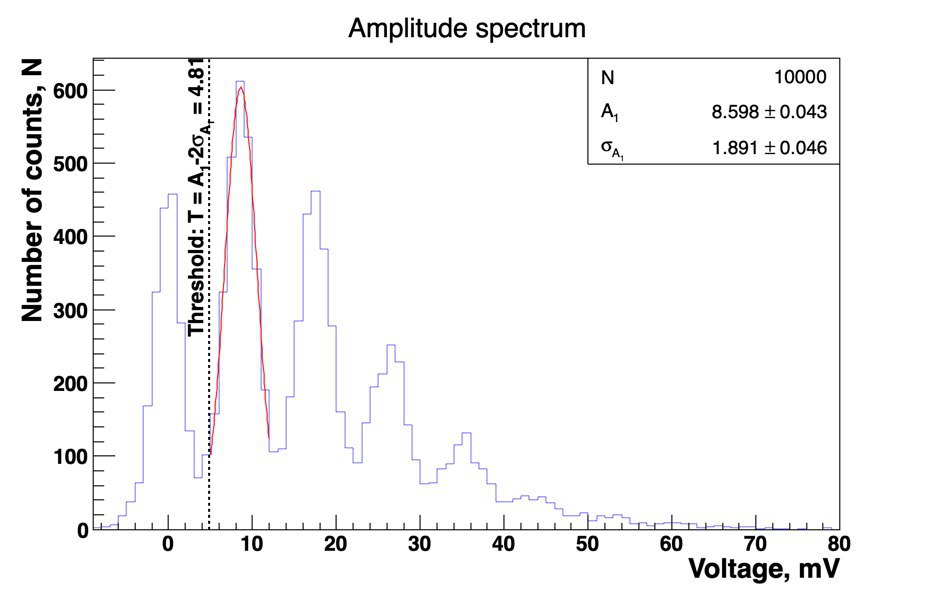}}
   
  \end{minipage}
  
  \hfill
  \caption{Typical charge and amplitude spectra of SiPMs: a - charge spectrum illustrate the pedestal method for evaluation of average number of photoelectrons. Blue area and $N_0$ - pedestal events on signal spectrum, gray area and $D_0$ - pedestal events on dark spectrum (incorporated picture); b - amplitude spectrum of SiPM and an illustration of the threshold extraction method.}
  \label{fig:fig_spectra}  
\end{figure}

By integrating waveforms within a fixed window (gate) we can extract charge\footnote{A SiPM's charge spectrum usually provides better single electron resolution compared to its amplitude spectrum (see~\cref{fig:fig_spectra}b). Usage of the charge spectra allows better extraction of pedestal events.} spectrum of  a SiPM. We acquire $N$-events for signal - (LED is ON), see~\cref{fig:fig_spectra}a, and $D$-events for dark (LED is OFF) spectra. Using the number of events in the pedestal for signal $N_0$ (blue area under the pedestal peak) and for dark $D_0$ spectra (gray area given in the same range) the estimation of the average number of photoelectrons $\hat{\mu}$ can be found \cite{anfimov} 
\begin{equation}
    \label{eq:mu_00}
   \hat{\mu} = - \ln{\left(\frac{N_0}{N}\cdot\frac{D}{D_0}\right)} = -\ln{\left(\frac{\hat{p}_{\xi_0}}{\hat{p}_{\lambda_0}}\right)} = \hat{\xi} - \hat{\lambda},
\end{equation}

where $\hat{p}_{\xi_0}$, $\hat{p}_{\lambda_0}$ - estimators of the pedestal probability in signal and dark  spectra,  and  $\hat{\xi} = -\ln(\hat{p}_{\xi_0})$, $\hat{\lambda} = -\ln(\hat{p}_{\lambda_0})$ - estimators of the average number of photoelectrons in signal and dark spectra, respectively.
Statistical error in this case could be evaluated \cite{anfimov}

\begin{equation}
    \label{eq:dispersion_mu_00_noise}
      \frac{\hat{\sigma}_{\hat{\mu}}}{\hat{\mu}} \approx \frac{1}{\sqrt{N}}\sqrt{\frac{e^{\hat{\xi}}+ e^{\hat{\lambda}}-2}{(\hat{\xi}-\hat{\lambda})^2}}.
\end{equation}

\begin{figure}[!h]
   \centering
  \includegraphics[width=350px]{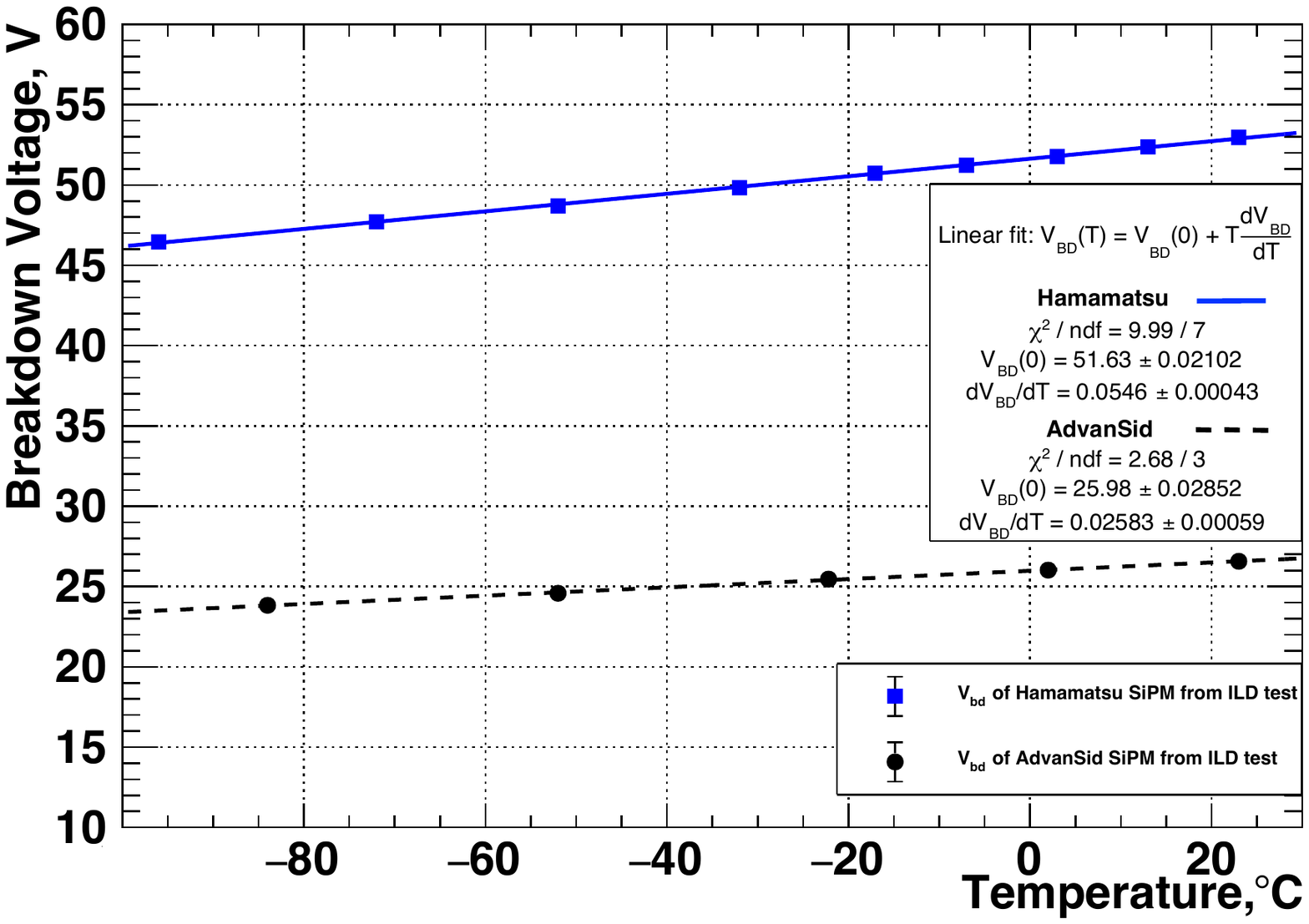} \\ 
   \caption{Breakdown voltages for Hamamatsu and AdvanSid SiPMs vs temperature using ILD test (see~\cref{eq:ILD}).}
  \label{fig:Vbd}
\end{figure}

 To compare parameters at different temperatures one should operate SiPM with similar overvoltage. A way to find the breakdown voltage is to analyze the IV-curve of a SiPM. There are several approaches of breakdown voltage extraction. In our study, we use the Inverse Logarithmic Derivative (ILD) test described in \cite{chmill}. The ILD for IV-curve reads

\begin{equation}
\label{eq:ILD}
   f(V) =  \left[\frac{1}{I}\cdot \frac{dI}{dV}\right]^{-1}
\end{equation}
and the minimum of the $f(V)$ yields the inflection point which defines the breakdown voltage. We also test the quadratic fit for AndvanSid SiPM which was used in the paper~\cite{italian}. Both approaches demonstrate good consistency of the breakdown voltage extraction (see~\cref{table:bd}).  Measured voltages for both SiPMs using the ILD test are shown in~\cref{fig:Vbd}.

\begin{table}[h]
\caption{\textbf{Breakdown voltages for the AdvanSid SiPM at different temperature points extracted by ILD test and quadratic fit}}
 \label{table:bd}
\centering
 \scalebox{0.8}{
 \begin{tabular}{c | l l l} 
 \hline
 \hline
\bf{SiPM type}  &  \multicolumn{3}{c}{\bf{Breakdown voltage $V_{bd}$ at different temperatures}} \\
 \hline
\multirow{2}{*}{ASD-NUV4S-P-4x4TD (ILD test)}  & 26.57 V[ $23~^o$C] & 26.02 V[ $2~^o$C] & 25.39 V[$-22~^o$C] \\
\cline{2-4}
  & 24.58 V[$-52~^o$C] & 23.83 V[$-84~^o$C]\\
\hline
\hline  
\multirow{2}{*}{ASD-NUV4S-P-4x4TD (quadratic fit)} & 26.55 V[ $23~^o$C] & 26.00 V[ $2~^o$C] & 25.39 V[$-22~^o$C] \\
\cline{2-4}
  & 24.57 V[$-52~^o$C] & 23.84 V[$-84~^o$C]\\
\hline
\hline
\end{tabular}}
\end{table}

 We evaluated the Dark Count Rate (DCR) of the SiPM using the following procedure. By processing raw waveforms we count a number of dark pulses (LED is OFF) that exceed some fixed threshold. We processed $w = 10000$ waveforms within $W = 1.024~\mu$s duration window\footnote{In our case, SiPM pulses $\sim100$~ns (FWHM) which can introduce a small fraction of pulses miscounting. We took it into account shrinking the window by the pulse width.}  (1024 samples at 1~GS/s sampling rate).  The overall counting window is $w \times W$. The DCR rate $R$ could be found from the number of dark pulses $N$ that have been counted in the overall window as
 \begin{equation}
    R = \frac{N}{w\times W} 
    \label{eq:dcr}
\end{equation}
 The number of the random pulses $N$ follows Poisson distribution\footnote{Standard deviation $\sigma_{N} = \sqrt{N}$} and a standard deviation of the DCR using \cref{eq:dcr} is
 
  \begin{equation}
     \sigma_R = \frac{\sqrt{N}}{w\times W}.
    \label{eq:err}
\end{equation}
 
 The threshold is obtained as an empirical compromise between efficiency and noise pick up. By measuring SiPM's amplitude spectrum (see.~\cref{fig:fig_spectra}b) we defined the optimal threshold\footnote{We assume signal and noise pulses are of similar amplitude.} $T$ as  
\begin{equation}
    \text{T} = A_1 - 2\sigma_{A_1},
    \label{eq:thr}
\end{equation}
where $A_1$ and $\sigma_{A_1}$ - the amplitude and its standard deviation of the single photoelectron signal, respectively.  Decreasing the threshold leads to significant noise pick up at a small gain (overvoltage). In our case (see \cref{eq:thr}) only $\sim$2,5~\% of SiPM's signals are miscounted in the dark rate evaluation procedure.\footnote{Assuming single pixel's amplitudes follow Normal distribution}
 
 \begin{figure}[ht]
  \begin{minipage}[ht]{0.5\linewidth}
  \label{fig:fig_pde_1}
  \centering
   \subcaptionbox{}
          {\includegraphics[width=\linewidth]{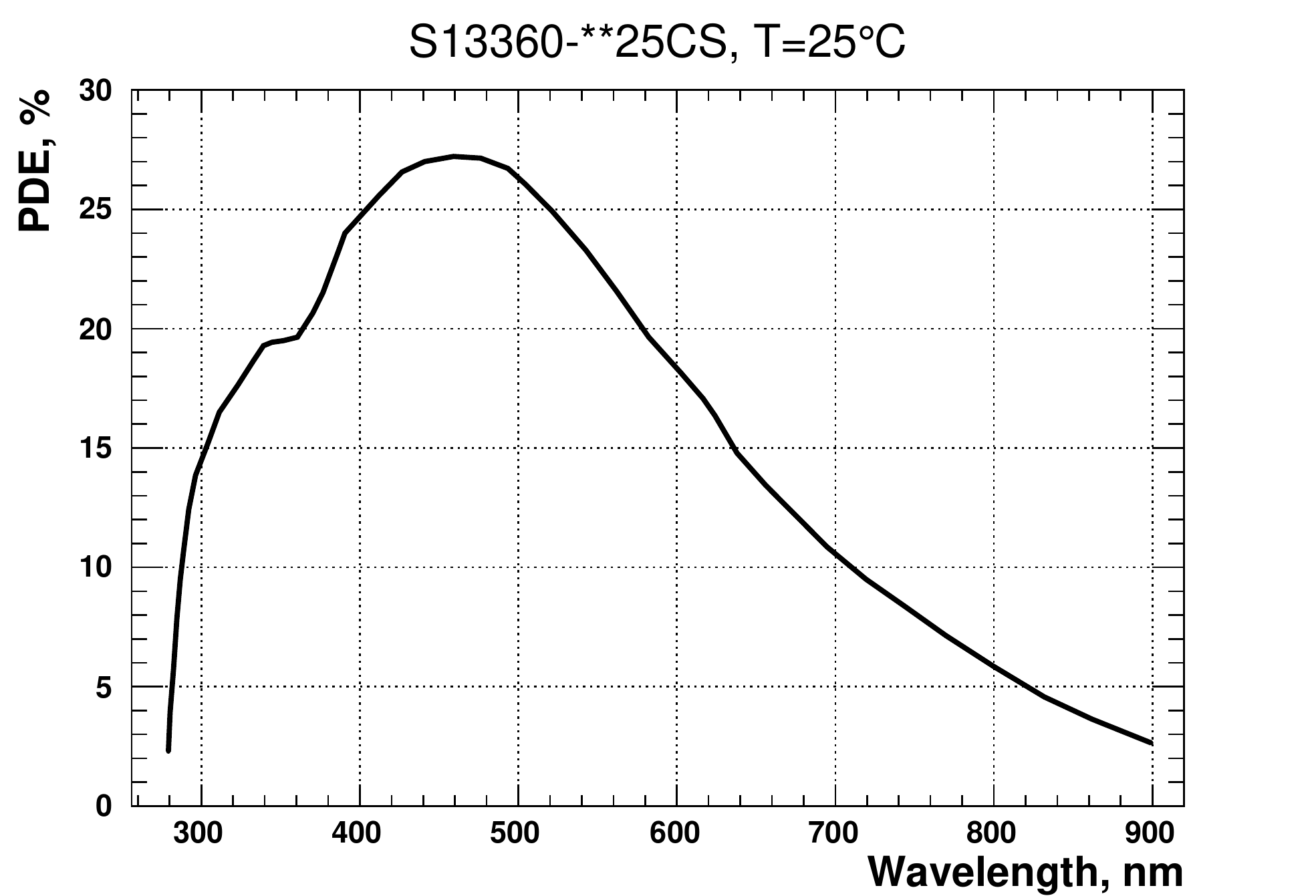}}
  \end{minipage}
  \hfill 
  \begin{minipage}[ht]{0.5\linewidth}
    \label{fig:fig_pde_2}
  \centering
   \subcaptionbox{}
          {\includegraphics[width=\linewidth]{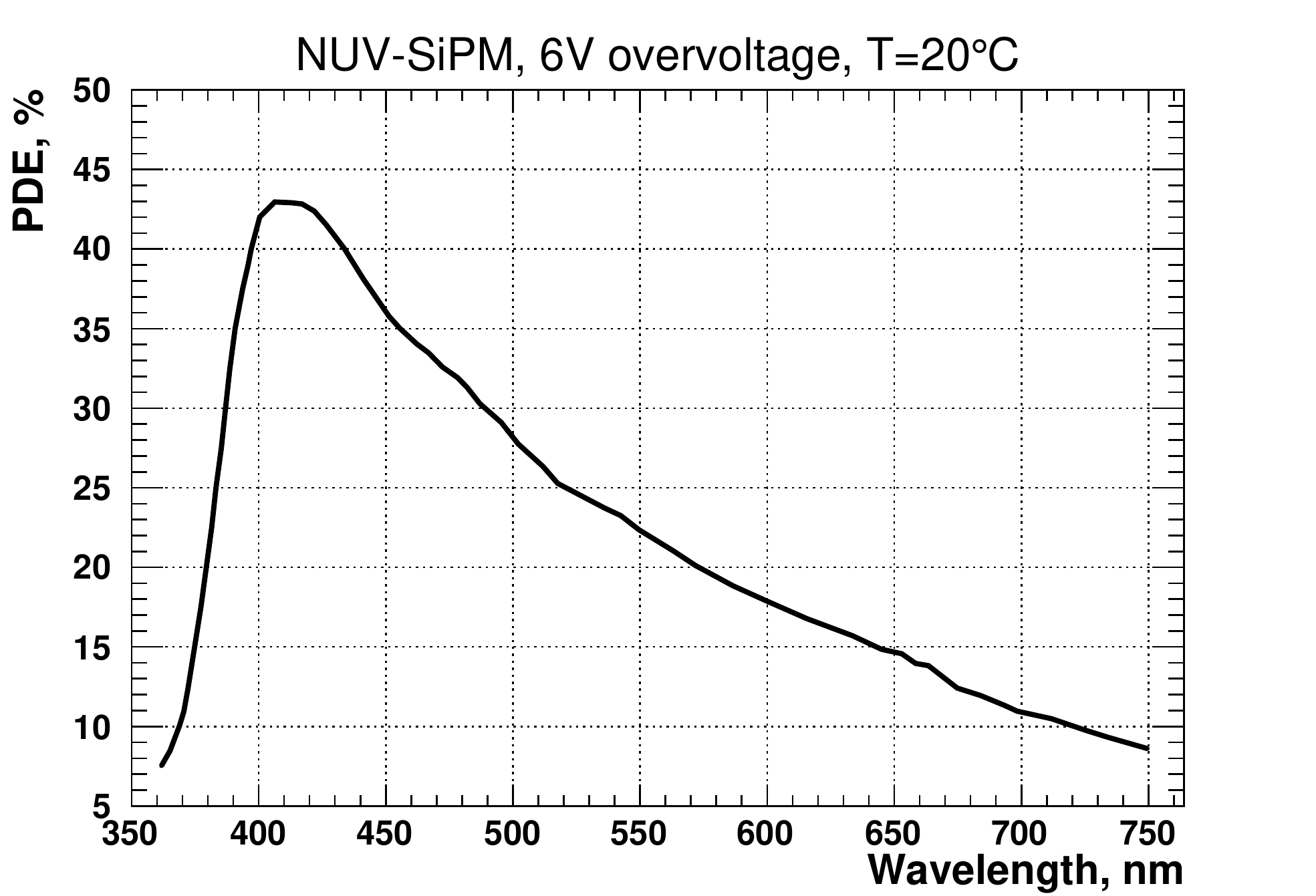}}
   
  \end{minipage}
  \hfill
  \caption{PDE as a function of wavelength: a -- for Hamamatsu SiPM at overvoltage OV = 5 V\cite{ham}, b -- for AdvanSid SiPM at OV = 6 V\cite{FBK}}
  \label{fig:fig_pde}  
\end{figure}

Measurements were performed for both SiPM samples at different temperatures and in the range of operating overvoltages. To present the PDE more illustratively we normalized measured relative PDE ($\hat{\mu}$) to 25~\% (see \cref{fig:fig_pde}) for Hamamatsu SiPM  ($V_{OV}=5.0~V$ at $T=23.0~^\circ$C) and to 42~\% \footnote{On \cref{fig:fig_pde}b the PDE is a bit higher than 40~\% at $V_{OV}$=6~V. For our sample, we were not able to reach $V_{OV}$ above 5.68~V because of high noise level. And we assumed the PDE=42~\% at this point.} for ASD-NUV4S-P-4x4TD ($V_{OV}=5.68~V$ at $T=23.0~^\circ$C).

\section{Results}
\label{sec:results}

The PDE and the DCR dependencies on overvoltage for two SiPMs were studied at the wide range of temperatures from $T=23.0~^\circ$C (room temperature) and even below the $T=-52.0~^\circ$C which is close to the TAO operational temperature.

\begin{figure}[htbp]
  \begin{minipage}[ht]{0.5\linewidth}
  \label{fig:fig_pde_1}
  \centering
   \subcaptionbox{}
          {\includegraphics[width=\linewidth]{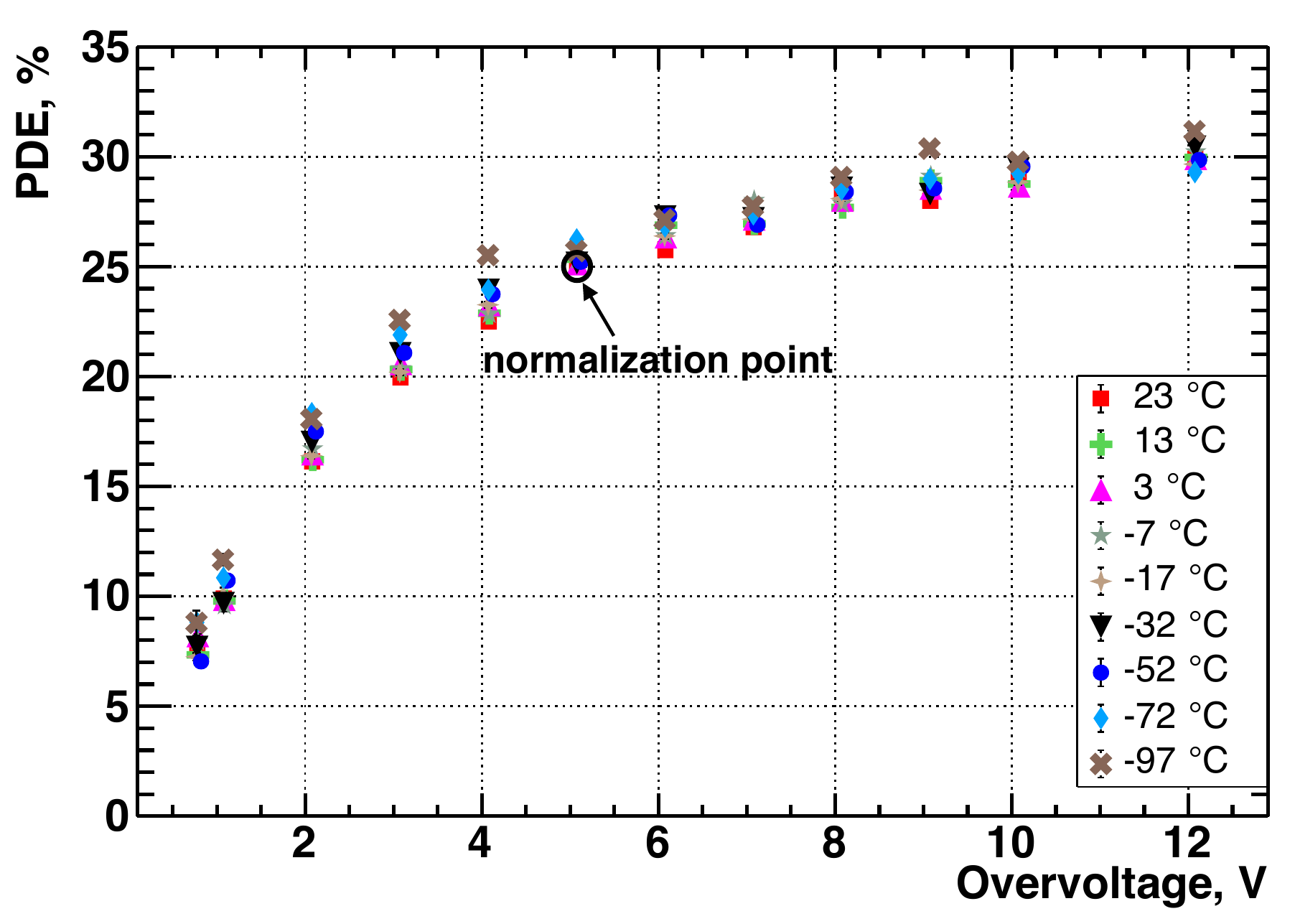}}
  \end{minipage}
  \hfill 
  \begin{minipage}[ht]{0.5\linewidth}
    \label{fig:fig_pde_2}
  \centering
   \subcaptionbox{}
          {\includegraphics[width=\linewidth]{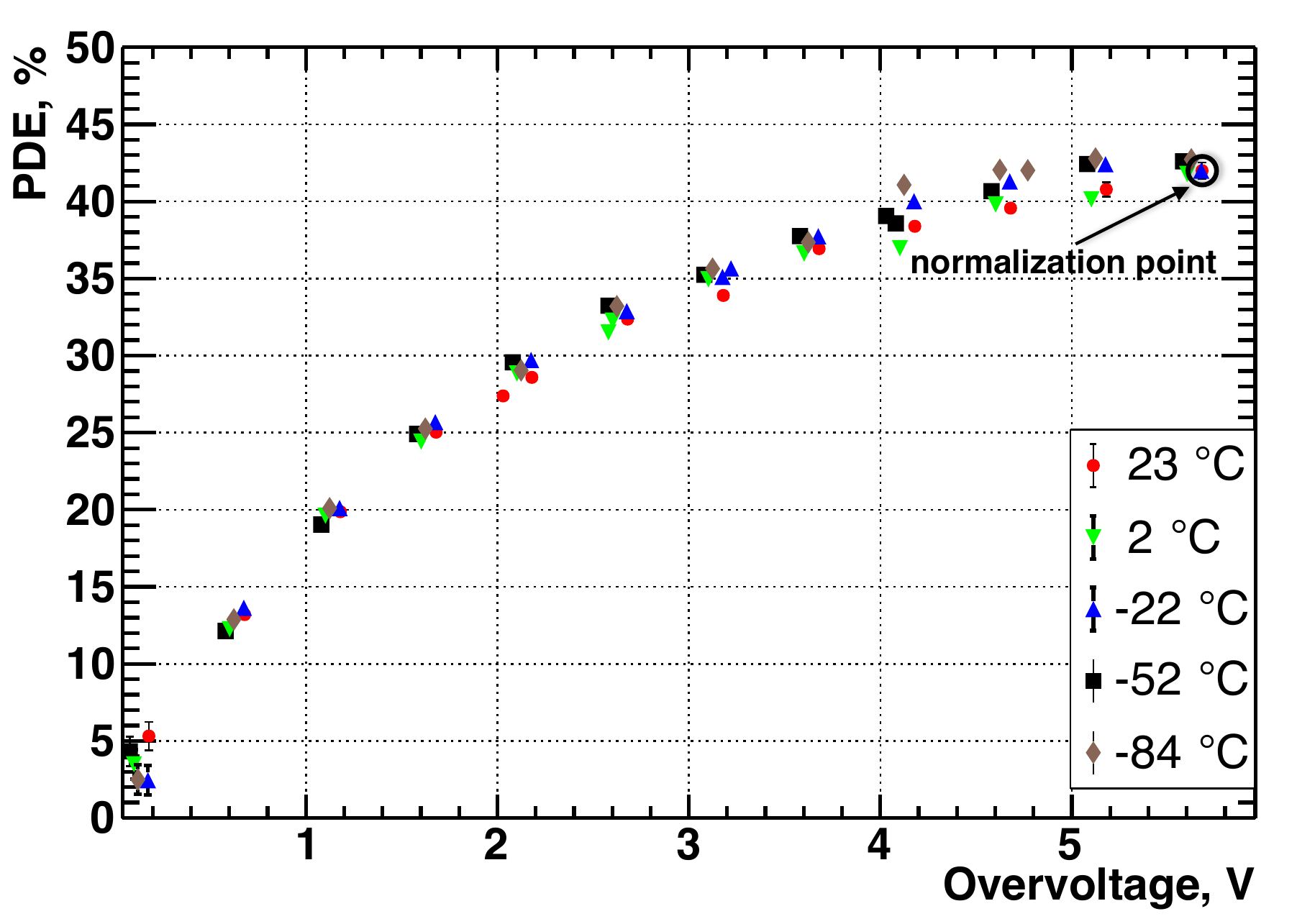}}
   
  \end{minipage}
  \hfill
  \caption{PDE as a function of overvoltage: a -- for Hamamatsu SiPM, b -- for AdvanSid SiPM} 
  \label{fig:fig_ov}  
\end{figure}

To compare the results of the PDE behavior for all temperatures we plot the PDE as a function of overvoltage (see.~\cref{fig:fig_ov}). One can see that there is no significant temperature dependence of the PDE for both samples. 

\begin{figure}[htbp]
  \begin{minipage}[ht]{1.0\linewidth}
  \label{fig:fig_dcr1}
  \centering
   \subcaptionbox{}
          {\includegraphics[width=\linewidth]{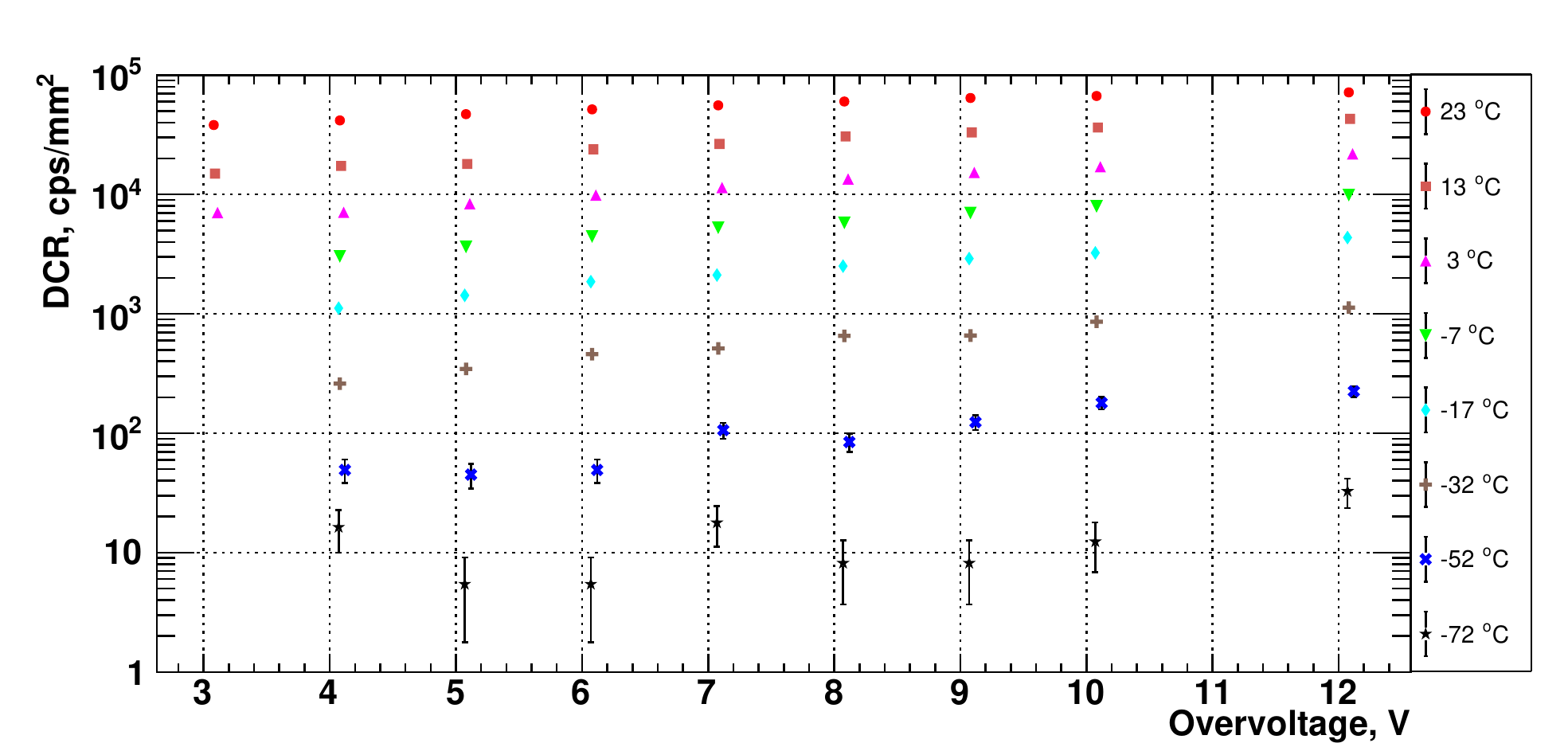}}
  \end{minipage}
  \hfill 
  \begin{minipage}[ht]{1.0\linewidth}
    \label{fig:fig_dcr2}
  \centering
   \subcaptionbox{}
          {\includegraphics[width=\linewidth]{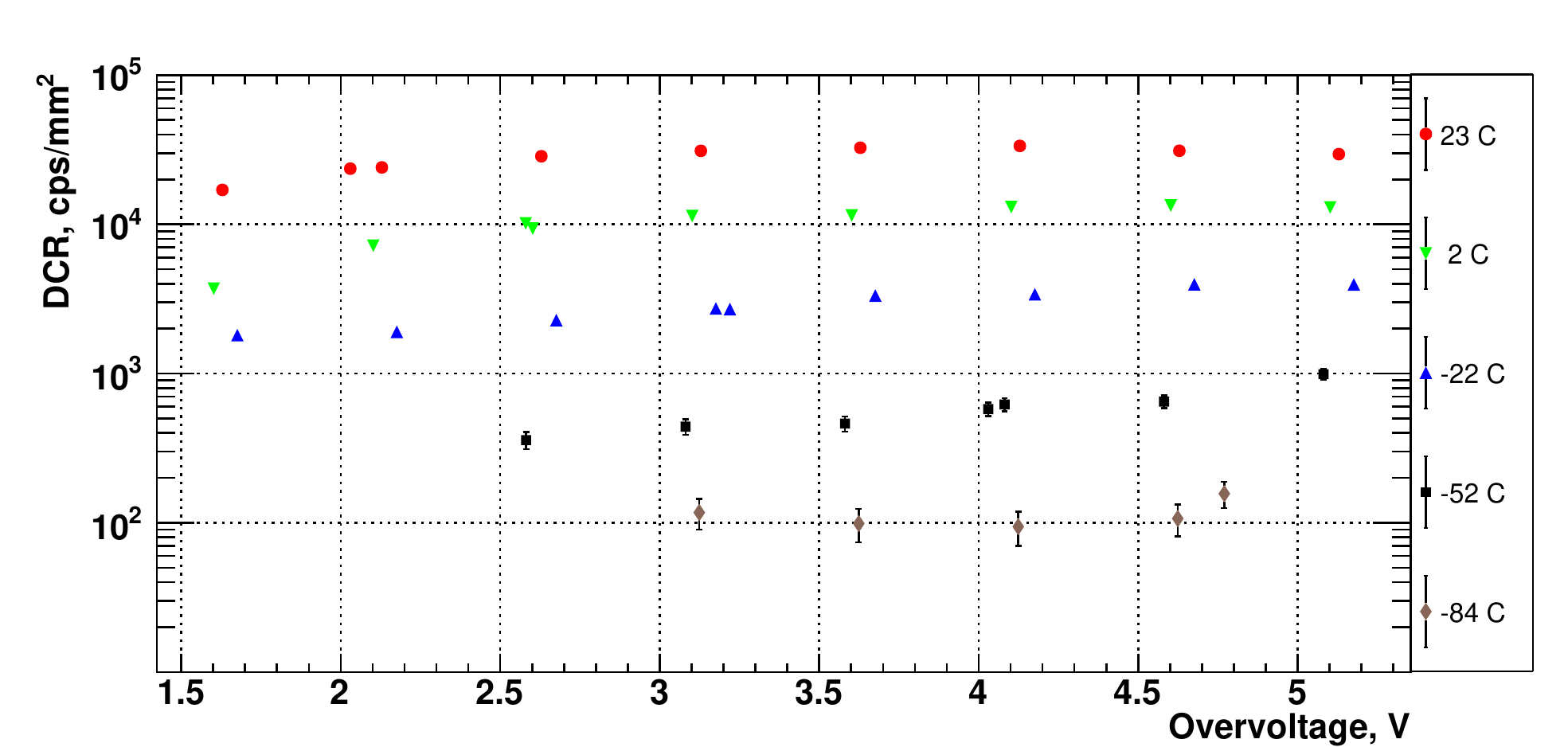}}
   
  \end{minipage}
  \hfill
  \caption{Evaluated DCR as a function of overvoltage: a -- for Hamamatsu SiPM, b -- for AdvanSid SiPM} 
  \label{fig:fig_dcr}  
\end{figure}

The DCR for SiPMs(see.~\cref{fig:fig_dcr}) decreases rapidly and reaches a drop by a factor of several orders of magnitude for low temperatures compared to room conditions. This fact is one of the major considerations for the SiPM usage at $T=-50~^\circ$C for the TAO detector. As can be seen from~\cref{fig:fig_dcr} for low overvoltages at negative temperatures we cannot obtain DCR values\footnote{Small number of dark pulses comparing to a noise pick up at a very low threshold}.

\section{Summary}
\label{sec:summary}
In general, the setup and technique demonstrated in this paper allows for measurements with different types of SiPMs in a wide range of temperatures from Liquid Nitrogen (77~K) up to the Room temperature.
We studied the Photon Detection Efficiency of SiPMs in a broad temperature range including $T = -52.0~^\circ$C ($\sim 220$~K) which is in the considered region of the TAO detector. As a result, two types of SiPMs - Hamamatsu S13360-6025CS and AdvanSID ASD-NUV4S-P-4x4TD have no significant difference in the PDE. These measurements do not exclude the behavior of the PDE in a broader temperature range and demonstrate an absence of difference for particular temperature points.\footnote{And incident light wavelength $\lambda=$425~nm.} 

The result presented in the study \cite{italian} that demonstrates the PDE drop in the temperature region around $T\sim$ 200~K to 220~K have not been found for our SiPM samples. Therefore, the SiPM usage in the TAO experiment is relevant and might have advantages comparing to the conventional PMTs. The choice of the proper type of SiPM for TAO requires a complex study of the SiPM's parameters at the TAO operational temperature.

\section{Acknowledgments}
The authors are indebted to Dr. Dmitriy Naumov, prof. Alexander~Olshevskiy and Christopher Kullenberg from the Joint Institute for Nuclear Research for their great help in the preparation of this paper. 
\label{sec:acknowledgments}
\bibliography{manuscript}
\end{document}